\documentclass[%
aip,
superscriptaddress,
amsmath,amssymb,
reprint,%
]{revtex4-1}

\usepackage{graphicx}
\usepackage{dcolumn}
\usepackage{bm}

\usepackage[utf8]{inputenc}
\usepackage[T1]{fontenc}
\usepackage{mathptmx}
\usepackage{siunitx}
\DeclareSIUnit\gauss{G}
\DeclareSIUnit[number-unit-product = {}]{\inchQ}{\textquotedbl}

\usepackage{braket}
\usepackage{amsmath}
\usepackage{physics}
\usepackage{tikz}

\begin{document}
	
	\preprint{AIP/123-QED}
	
	\title{A compact and fast magnetic coil for the interaction manipulation of quantum gases with Feshbach resonances}
	
	\author{A. Kell}
	\author{M. Link}
	\author{M. Breyer}
	\author{A. Hoffmann}
	\author{M. Köhl}
	\email{michael.koehl@uni-bonn.de}
	\affiliation{Physikalisches Institut, University of Bonn, Wegelerstrasse 8, 53115 Bonn, Germany}
	
	\author{K. Gao}
	\thanks{Author to whom correspondence should be addressed: kgao@ruc.edu.cn}
	\affiliation{Physikalisches Institut, University of Bonn, Wegelerstrasse 8, 53115 Bonn, Germany}
	\affiliation{Department of Physics, Renmin University of China, 100872 Beijing, China}

	%

	\date{\today}
	
	\begin{abstract}
		Cold atom experiments commonly use broad magnetic Feshbach resonances to manipulate the interaction between atoms. In order to induce quantum dynamics by a  change of the interaction strength, rapid ($\sim\SI{}{\us}$) magnetic field changes over several tens of Gauss are required. Here we present a compact design of a coil and its control circuit for a change of the magnetic field up to \SI{36}{\gauss} in \SI{3}{\us}. The setup comprises two concentric solenoids with minimal space requirements, which can be readily added to existing apparatuses. This design makes the observation of non-equilibrium physics with broad Feshbach resonances accessible.
	\end{abstract}
	
	\maketitle

	\section{\label{sec:level1}INTRODUCTION}
	
	Ultra-cold quantum gases with tunable interactions have been widely used to investigate strongly correlated matter in the last decade\cite{Bloch08}. Magnetic Feshbach resonances change the coupling of different scattering channels in the low temperature regime, and control the interparticle interaction in quantum gases\cite{Ketterle98}. Making use of this extraordinary tunability requires a change of magnetic field on the order of tens of Gauss for a broad Feshbach resonance to accurately control the interaction strength\cite{Chin10}. Experiments investigating non-equilibrium phenomena in quantum gases take advantage of Feshbach resonances to induce fast dynamics beyond current theoretic understanding\cite{Makotyn14,Eigen18}. However, it is challenging to vary the magnetic field with significant amplitude on time scales set by fundamental energy scales, e.g. the Fermi energy, to a few microseconds\cite{Langen15,Yuzbashyan15,Polkovnikov11}.
	
	Most experiments use pairs of coils centered around the atom position to generate a near-homogeneous magnetic field and tune the interaction of the atoms. Because of the high magnetic field, the coils need to be very close to the atoms and efficiently cooled\cite{Streed06,Sabulsky13}. However, high magnetic field and spatial restriction lead to multi-winding and large size of the coils, which result in large self-inductance and mutual inductance between the coils and experimental setup. Additionally, to facilitate imaging systems with large numerical aperture, the atom position is often not in the center of the vacuum system, but close to a viewport. This further increases the required distance between the pair of coils. 
	
	The difficulty to change the magnetic field fast enough results from the quick change of the currents in the coils and the induced eddy currents in the nearby materials, e.g. metallic vacuum chambers or gaskets. Some recent experiments realise fast change of magnetic field using auxiliary coils\cite{Eigen19} and overshooting compensation fields\cite{Dedman01,Olsen08} to overcome these limitations. However, they only address narrow Feshbach resonances and cause small changes of the scattering length\cite{Chin10}. It is still difficult to access a fast change of scattering length with large amplitude in broad Feshbach resonances.
	
	Here we demonstrate a compact design of two concentric coils to change the magnetic field by up to \SI{36}{\gauss} in \SI{3}{\us}, which is fast enough to access non-equilibrium physics with most Feshbach resonances in alkali atomic species. Because the coils have different sizes and opposite current circulation, the magnetic field gradients cancel while the offset field is reasonably high. The small size and in-serial configuration of these coils lead to very low self-inductance. Meanwhile, this compact design of two coils on only one side of the vacuum chamber confines the magnetic field to a relatively small space. Therefore it is possible to switch off both the current and the magnetic field quickly. The 3D-printed plastic mount used here not only avoids any eddy current, but also makes it very flexible to construct. This design can easily be adapted to different experiments to access non-equilibrium physics of quantum gases with most broad Feshbach resonances\cite{Chin10}. It might find more applications in other apparatuses requiring fast switching or modulating of the magnetic fields, including quantum simulation\cite{Bloch08}, quantum sensing\cite{Degen17}, magnetic resonance imaging\cite{jensen1987}, electron microscopes\cite{golladay1988} and even biomagnetic experiments\cite{platzek87}.
	
	\section{\label{sec:level1}FAST COIL}
	\subsubsection{\label{sec:level1}DESIGN OVERVIEW}
	
	Our design to realise a fast change of magnetic field addresses three considerations: (1) a magnetic coil having small enough self-inductance and providing large field strength, (2) the coil coupling only weakly to the metallic vacuum chamber and other coils, and (3) a control circuit switching the current quickly.
	
	\begin{figure}
		\includegraphics[scale=0.35]{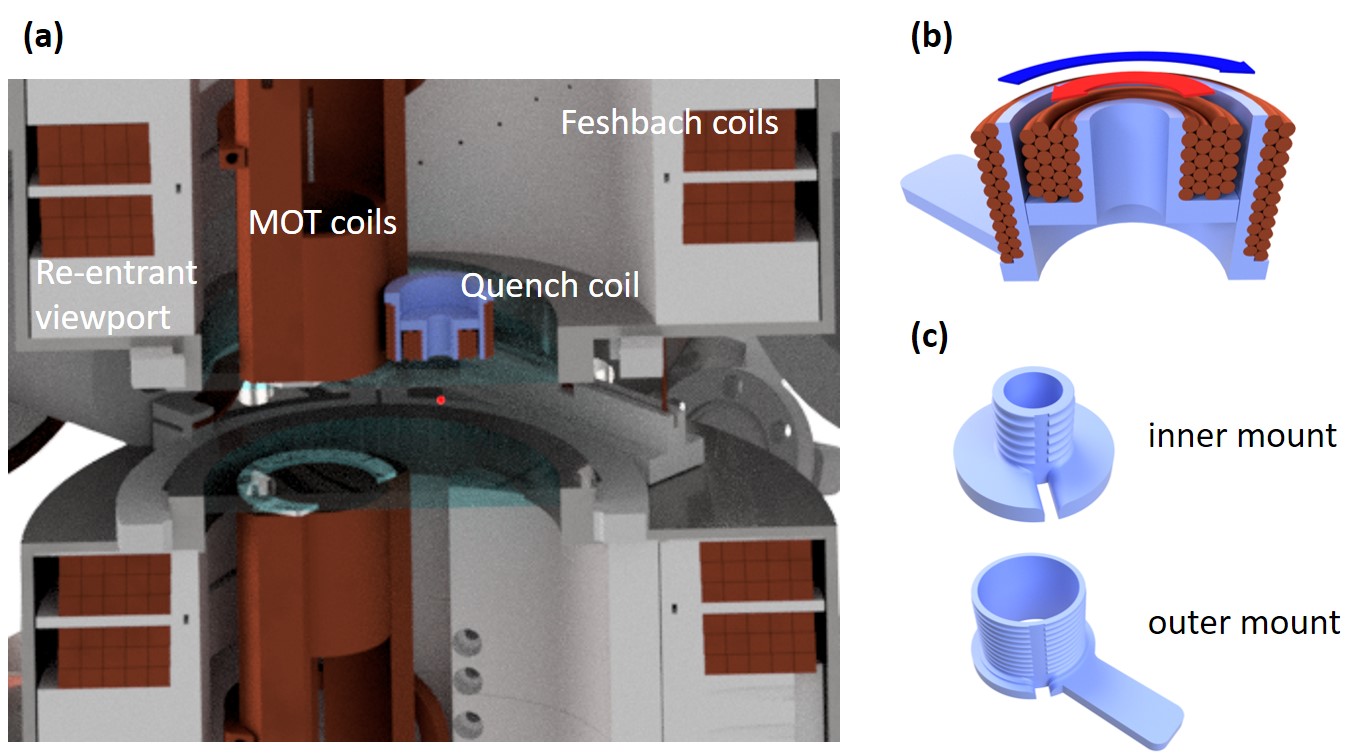}
		\caption{\label{fig:1} Experimental setup and coils. (a) Schematic drawing of the metallic vacuum chamber and magnetic field coils. The red dot indicates the atoms in the chamber. (b) Design of the fast coil. The blue and red arrows show the current through the coils. (c) CAD drawing of the 3D printed mount of the coils.}
	\end{figure}
	
	As shown in Fig. 1(a), the main parts of our experimental setup include an \SI{8}{\inchQ} extended spherical octagon stainless steel vacuum chamber (Kimball Physics MCF800-ExtOct-G2C8A16), two re-entrant viewports (UKAEA) with \SI{65}{\milli\metre} diameter windows, a pair of MOT coils and multi-layer copper-tube Feshbach coils with near-Helmholtz configuration\cite{Harrison17}. The ${^6}$Li atoms are trapped \SI{3}{\milli\metre} below the \SI{3.5}{\milli\metre} thick top window and the fast coil stays about \SI{0.5}{\milli\metre} above it. The Feshbach coils provide up to \SI{1000}{\gauss} magnetic offset field, while the fast coil generates an extra \SI{36}{\gauss} for a fast change of the total magnetic field.\\
	
	The fast coil consists of two small concentric coils of different size. Because of the geometry of the coils and the opposite current circulation, the field gradients of the coils cancel at a certain position $z_0$ along the symmetry axis. The magnetic field on the $z$ axis near the position $z_0$ can be calculated from the Biot-Savart law
	\begin{equation}
	B(z)= B_1(z_0)-B_2(z_0)+O((z-z_0)^2)
	\end{equation}
	where $B_1(z_0)$ and $B_2(z_0)$ are the field contribution from the coils with opposite current circulation, and $z_0$ is determined by the geometry and relative position of the two coils. 
	
	To suppress the eddy currents in the metallic chamber and gaskets close to the fast coil, the mutual inductance between the fast coil and the experimental setup has to be small. The coils are connected in series and the currents are running in opposite direction so that magnetic field is confined in a very limited space. The mutual inductance between the coils and the chamber is significantly reduced. This configuration also guarantees the stability of the magnetic field since the field gradients are determined by the geometry of the coils and always precisely cancel at $z_0$. Meanwhile, we mount the coil in a 3D-printed plastic mount in order to avoid eddy currents.
	
	The coil is driven by a programmable power supply in constant current mode and the current is controlled by an external analog signal. The fast switch is realized by a power MOSFET switching off the current of the coil and dissipating the magnetic energy with an RC snubber circuit.
	
	\subsubsection{\label{sec:level1}FAST COIL}
	
	As shown in Fig. 1(b), the fast coil consists of two concentric coils of different sizes to provide a reasonable magnetic field and fast switching. We use a polyimide coated copper wire with diameter \SI{1}{\milli\metre} here. The outer coil consists of two layers with 22 windings and is \SI{22}{\milli\metre} in diameter. The inner has 22 windings in four layers and \SI{6}{\milli\metre} in diameter. Both coils are wound on the 3D-printed plastic mounts and connected in series. In our design, the field gradients cancellation occurs at $z_0=\SI{9}{\milli\metre}$ along the axial direction, that is \SI{8}{\milli\metre} from the bottom surface of the coil. The magnetic field at this point is \SI{36}{\gauss} at \SI{40}{\ampere}. The magnetic field distribution along the $z$ direction is shown in Fig. 2(a), along the radial direction in Fig. 2(b). The total resistance of the coil is $\SI{0.1}{\Omega}$, and total inductance is calculated to be \SI{6}{\micro\henry}.
	
	\begin{figure}
		\includegraphics[scale=0.3]{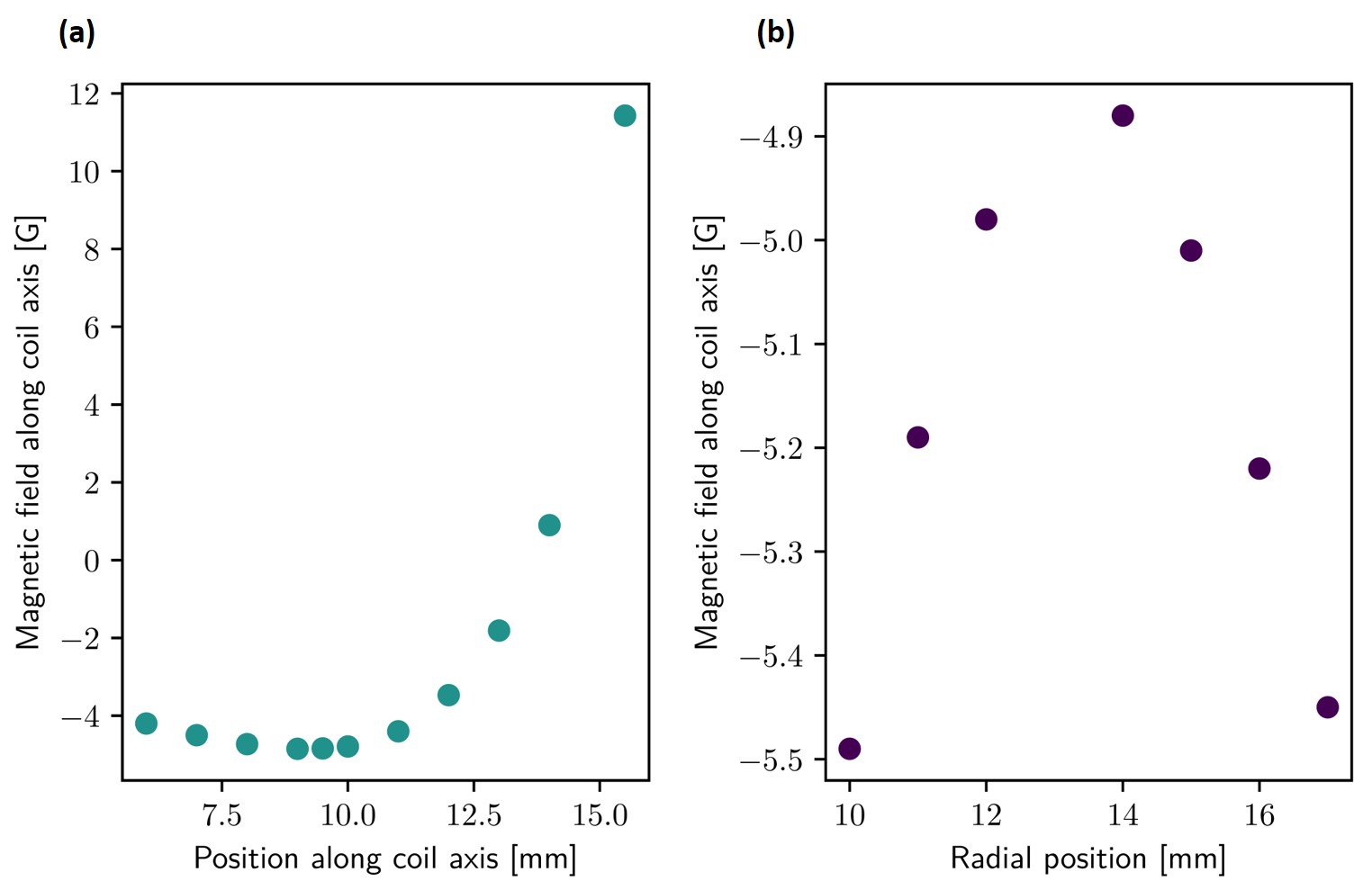}
		\caption{\label{fig:2} Profile of the magnetic field of the fast coil. (a) Magnetic field along $z$ direction. (b) Magnetic field in radial directions of the coil at $z_0=\SI{9}{\milli\metre}$. All the fields are measured at a current of 5A through the coil. The magnetic field at 40A is verified to be \SI{36}{\gauss} by Zeeman spectroscopy with atoms.}
	\end{figure}
	
	\subsubsection{\label{sec:level1}3D PRINTED MOUNT AND COIL CONSTRUCTION}
	The mount of the coil is directly 3D printed with the fused filament fabrication method and polylactide (PLA) is used. This material is chosen because of its easiness to print and low cost, so that different designs could be easily tried and practically optimized. The 3D drawing of the mounts are shown in Fig. 1(c). The round groove on the wall of the mounts is printed to easily position the wire. With the printed mounts, both coils are first wound separately. Some epoxy is used to fix the outer layers of both coils. Then they are aligned, glued and connected together. The mounted coil is held by a plastic post and a 3D translation stage. A low duty cycle of \SI{2}{\percent} is chosen to avoid overheating of the coil. It is high enough for a typical cycle time of \SI{25}{\second} when the coil is on for \SI{0.5}{\second} for the atoms to thermalize. Two temperature sensors (Pt100) are glued on the outer coil to monitor the temperature for the external interlock circuit. Since the coil is not water cooled, the temperature reaches at most \SI{25}{\celsius} above the room temperature during our typical experimental cycle.
	
	\subsubsection{\label{sec:level1}ALIGNMENT}
	The compact design of this coil does not provide a near-homogeneous magnetic field over a large volume, therefore the alignment has to be done in two steps. For the coarse alignment, the Zeeman shift of the optical transition of thermal ${^6}$Li atoms is used to determine the maximum of the magnetic field with a precision of \SI{0.1}{\gauss} in z direction when the position of the coil is changed by a 3D translation stage. At the same time, the displacement of the atomic cloud in x and y directions can be used to optimize the radial directions. To finely optimize the position, the RF transition frequency of thermal ${^6}$Li atoms is used to measure the magnetic field with a precision of \SI{10}{\milli\gauss}. The fast coil can be aligned so that the gradient is minimized to the order of magnitude of \SI{1}{\gauss\per\centi\metre} or less and has negligible effect on the original trap. In our experiments with a ${^6}$Li Fermi gas, collective modes of the atomic cloud are observed when we introduce a small magnetic gradient by slightly moving the fast coil from the optimal position. In principle, a small magnetic field gradient can also be used to accelerate the evaporation cooling of quantum gases in optical dipole traps \cite{Hung08}.
	
	\section{FAST SWITCHING OF THE MAGENTIC FIELD}
	\subsubsection{\label{sec:level1}CONTROL CIRCUIT}
	
	The current through the coil is provided by a programmable power supply (Delta Elektronika S280) in constant current mode. The main functionality of the control circuit is to switch off the current through the coil quickly, as to change the magnetic field. As shown in Fig. 3, the circuit schematic can be divided into two sections: the current switch and the switch signal processor. 
	
	The heart of the circuit is the current switch. To shut off the current through the coil, a high-speed power MOSFET driver (Microchip Technology MCP1407) and a power MOSFET (IXYS  IXFN140N20P) are chosen. The high peak output current of \SI{6}{\ampere} of this driver enables switch time down to \SI{30}{\nano\second}. The drain-source on-state resistance of the power MOSFET is less than $\SI{18}{\milli\Omega}$ so that the total DC resistance of coil, wiring and MOSFET is less than $\SI{150}{\milli\Omega}$. The current is only limited by the power supply to \SI{40}{\ampere}.
	
	\begin{figure}
		\includegraphics[scale=1]{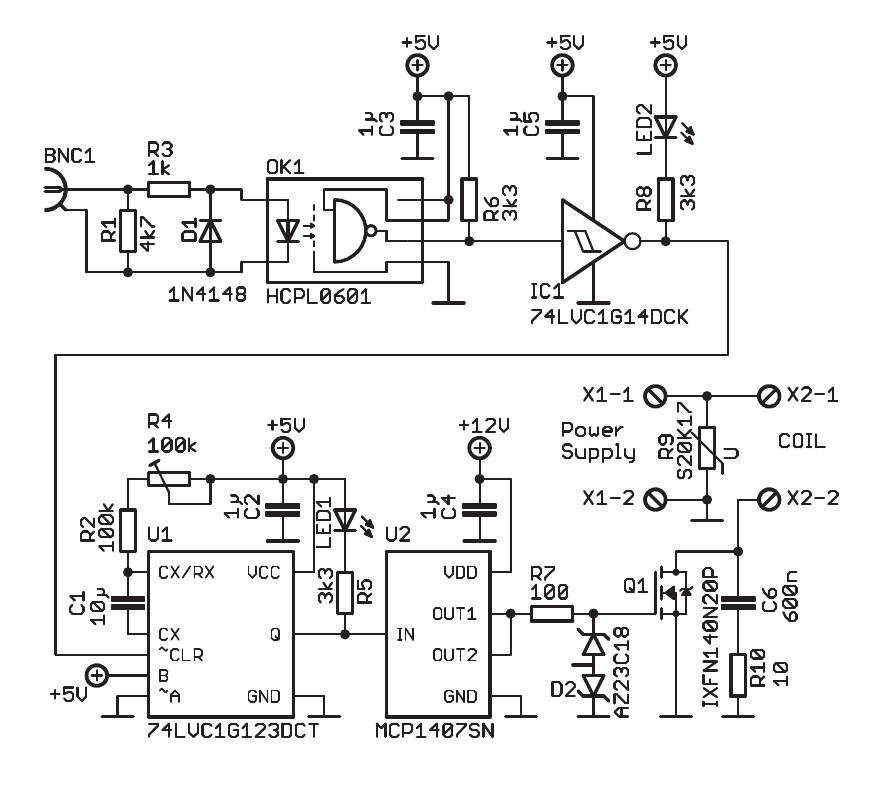}
		\caption{\label{fig:3} Schematic of the control circuit of the fast coil.}
	\end{figure}
	
	The switching of the current is also affected by the coil and nearby components. When the current from the power supply is cut off by the power MOSFET, the current in the coil is driven to zero by dissipating the magnetic field energy with an RC snubber circuit including a power resistor and high voltage capacitors. Here a carbon composition resistor with no inductance (RC07GF100J) and three parallel connected 200V capacitors (SMD 1812) are chosen. The snubber circuit is optimised for fast switching by LTspice simulations. When the MOSFET turns off, the current can flow to the AC coupled snubber resistor, which dissipates the magnetic energy of the coil. The lead of the fast coil to the control circuit is less than \SI{15}{\centi\metre} and the power cables are connected directly to the MOSFET so that the switching performance is not compromised.
	
	To process the TTL switch signal from an external system, a switch signal processor circuit is used here. It includes a digital opto-coupler (HCPL-0601), a single Schmitt-trigger inverter (Texas Instruments SN74LVC1G14) and a single retriggerable monostable multivibrator with Schmitt-trigger inputs (Texas Instrumts SN74LVC1G123). The optocoupler isolates the input from our experiment control. The inverter corrects the signal polarity, and a retriggerable monostable multivibrator limits the on-time of the MOSFET to about \SI{1}{\second} to avoid overheating of the coil.
	
	\subsubsection{\label{sec:level1}SWITCHING CHARACTERIZATION}
	
	To characterize the performance of the fast coil, especially the fast change of the magnetic field, we would like to directly measure the magnetic field at the atom position. A pick-up coil as a sensor is used to measure the changing magnetic flux close to the fast coil. A \SI{15}{\milli\metre}-diameter coil with two turns is used to detect an induced signal to an oscilloscope as a measurement of the changing flux. The temporal change of the magnetic field is checked in two different ways: Firstly, the pick-up signal is measured when the coil is switched off without metallic parts, including a CF-40 vacuum flange, a CF-40 gasket and a prototype of our Feshbach coils, as shown in Fig. 4. It is seen that the magnetic field is turned off (the integrated pick-up signal, from 90\% to 10\%) in about \SI{3}{\us}, which is shorter than a typical time scale set by the Fermi energy. The switching performance is slightly degraded with the existence of the above metallic parts around the fast coil, due to the small mutual inductance of the fast coil and the nearby metallic objects. Secondly, when the coil is placed above the top window of the re-entrant viewports in our experimental setup, the pick-up signal is also measured. It shows the same switching speed, which indicates the mutual inductance between the fast coil and the experimental setup is negligible. In addition, the fast coil is also tested with current modulation up to \SI{100}{\kilo\hertz}, which could be used to measure the excitation spectrum in quantum gases\cite{Greiner05, Behrle18}.
	
	\begin{figure}
		\includegraphics[scale=0.6]{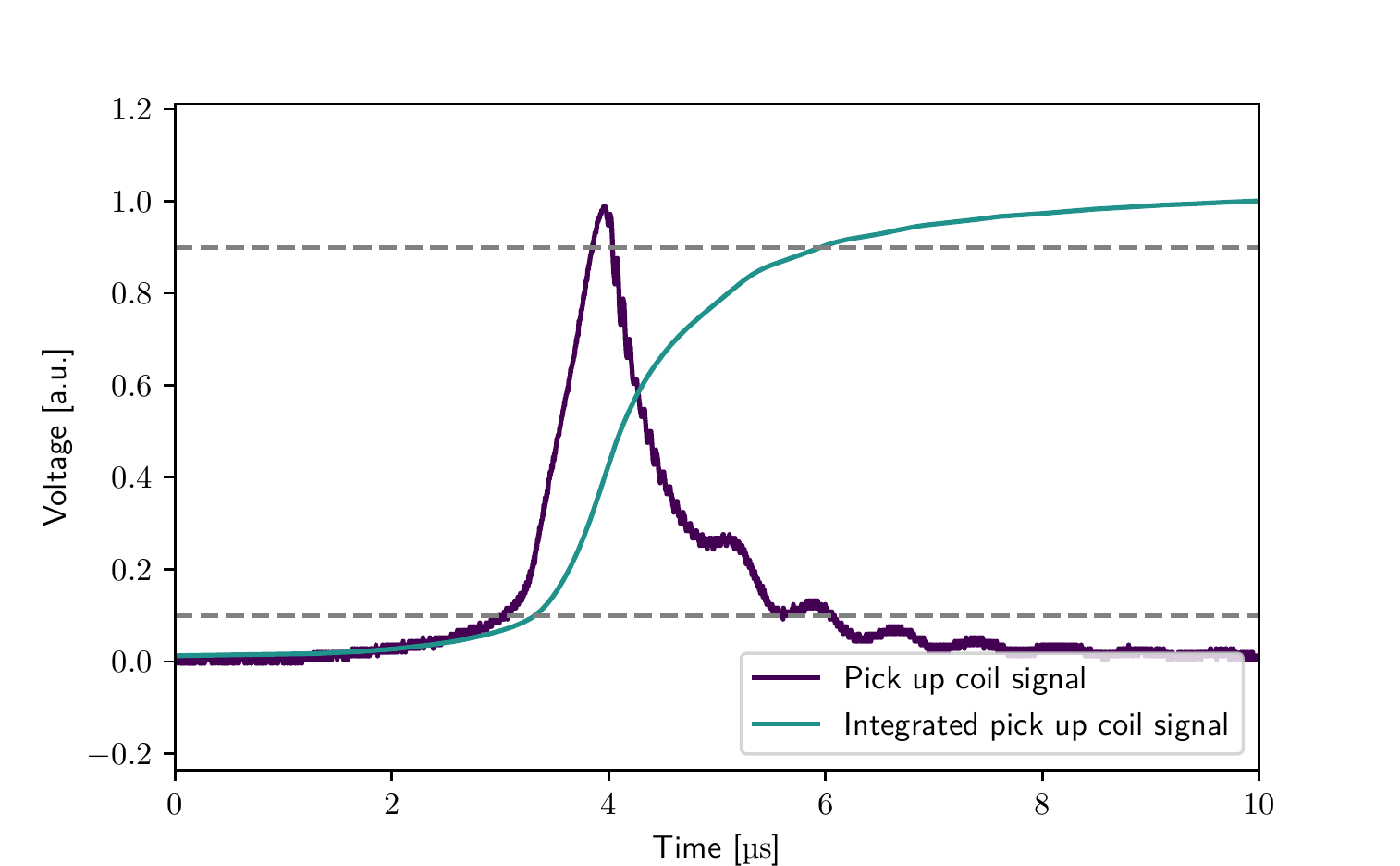}
		\caption{\label{fig:4} Fast change of the magnetic field of the fast coil. The pick-up signal (purple) and its integrated signal (blue) during the magnetic field switching, the dashed lines indicate 10\% and 90\% voltage levels of the integrated signal. The time of the magetic field change is about \SI{3}{\us}. }
	\end{figure}
	
	\section{CONCLUSION AND OUTLOOK}
	We have presented a compact coil design and its control circuit for a fast change of magnetic field up to \SI{36}{\gauss} within \SI{3}{\us} for broad Feshbach resonances. This design enables dynamic manipulation of the interaction in most fundamental time scales and allows for exploration of non-equilibrium physics in quantum gases with broad Feshbach resonances\cite{Yuzbashyan15, Langen15}. In our experiment, it enables a quantum quench of the interaction strength $1/k_Fa$ of a 1-3 mixture of ${^6}$Li by up to 0.6 at unitarity. Its compactness makes it useful for setups with limited space, especially when the atoms are not in the center of the vacuum chamber. In future, a high magnetic gradient could also be realized by a similar design with cancelling offset fields, which is useful for preparation of neutral atoms in quantum gas microscopes\cite{Sherson10} and atom-based quantum sensors\cite{Degen17}.
	
	\section*{acknowledgments}
	This work has been supported by BCGS, the Alexander-von-Humboldt Stiftung, ERC (grant number 616082), DFG (SFB/TR 185 project B4), and under Germany's Excellence Strategy-Cluster of Excellence Matter and Light for Quantum Computing (ML4Q) EXC 2004/1-390534769.
	
	\section*{DATA AVAILABILITY}
	The data that support the findings of this study are available from the corresponding author
	upon reasonable request. 
	
	\nocite{*}
    \bibliography{bitex}

\end{document}